\documentclass[11pt]{article}

\usepackage[utf8]{inputenc}
\usepackage[T1]{fontenc}
\usepackage{lmodern}
\usepackage{microtype}
\usepackage[margin=1in]{geometry}
\usepackage{graphicx}
\usepackage{booktabs}
\usepackage{amsmath}
\usepackage{amssymb}
\usepackage{authblk}
\usepackage[numbers,sort&compress]{natbib}
\usepackage{hyperref}
\usepackage{url}
\usepackage{orcidlink}

\hypersetup{
  colorlinks=true,
  linkcolor=blue,
  citecolor=blue,
  urlcolor=blue,
}

\title{Quieting the Cobwebs: Browser Interaction for Visual Floaters}

\author[1]{Kenneth Ge\,\orcidlink{0009-0000-5044-4433}\thanks{kennyge@assistivity.org}}
\author[2]{Shikhar Ahuja\thanks{ahujashikhar314@gmail.com}}
\author[3]{Jinglin Li\thanks{jinglinl928@gmail.com}}

\affil[1]{Assistivity, Edgemont, NY, USA}
\affil[2]{Georgia Institute of Technology, Atlanta, GA, USA}
\affil[3]{Independent Researcher, San Francisco, CA, USA}

\date{\today}

\begin{document}

\maketitle

\begin{abstract}
Floaters, cobweb-like shadows that move around a person's visual field, impair vision for nearly 33\% of the population, yet have limited treatment options. Floaters especially harm screen use, since they reduce contrast, introduce clutter, and add moving distractions. While existing high-contrast tools offer some help, few address the motion that makes screen use with floaters uniquely difficult. In this paper, we build a floater simulation inspired by the physics of the eye, use it to quantitatively assess text readability at varying levels of motion, and build a novel web extension that minimizes eye movement, maximizing the signal-to-noise ratio of performing browser tasks. Importantly, our tool works not only for text, but for all UI elements, requiring no modifications to existing websites.
\end{abstract}


\section{Introduction}
You move your eyes to the left. Cobweb-like lines slowly drift to follow. You move your eyes back to the right. The lines twist and swirl as they move toward the right edge of your vision. If you can relate to this, you may have a phenomenon known colloquially as visual floaters. Formally, these ``vitreous opacities'' are bundles of collagen in the liquid portion of the eye that cast translucent gray shadows onto the retina \cite{noauthor_vitreous_nodate}. While floaters are sometimes caused by sight-threatening conditions, more often they are associated with nearsightedness \cite{noauthor_vitreous_nodate}. Indeed, floaters affect an estimated 33\% of smartphone users worldwide \cite{webb_prevalence_2013}.

Unfortunately, treatment options for floaters are limited. Floaters typically remain in the eye permanently \cite{noauthor_what_nodate}. While technologies like Ortho-K delay the progression of myopia \cite{charm_high_2013}, they don't clear away floater debris in the vitreous. Barring folk remedies like pineapple juice \cite{noauthor_ive_nodate}, the only treatments that remove them, YAG laser vitreolysis and vitrectomy \cite{noauthor_efficacy_nodate, sebag_vitrectomy_2014}, carry a high risk of complications including retina and cataract damage.

Most people adapt to floaters. But for individuals with higher floater severity, or in professions that require clear, high-contrast vision like radiology \cite{the_floater_doctor__james_h_johnson_md_vitreous_2020}, they can prove distressing. Individuals with floaters were willing to trade 11\% of their remaining lifespan to permanently get rid of their floaters \cite{wagle_utility_2011}, and were more likely to have anxiety and depression \cite{gouliopoulos_association_2024}. Floaters affect individuals of all age groups, races, genders, and eye colors \cite{webb_prevalence_2013}, and do not always become less bothersome with time \cite{katsanos_safety_2020}.

Floaters harm screen use by degrading contrast sensitivity \cite{sebag_vitreous_2020}, blocking or obstructing text or other elements \cite{woudstra_de_jong_impact_2025}, graying colors \cite{admin_donna_2021}, and introducing visual clutter. Clutter is especially significant because it causally reduces reading \cite{woudstra_de_jong_impact_2025} and object recognition \cite{admin_donna_2021} speed by obstructing information flow in the visual cortex \cite{pelli_crowding_2007, legge_psychophysics_1985}, even for individuals with high visual acuity \cite{levi_crowding--essential_2008}. In a static context, the human visual system can fill in missing details \cite{noauthor_features_nodate, wagemans_century_2012}, adapt to global degradation \cite{sawides_vision_2011, webster_neural_2002}, and filter out non-moving visual stimuli \cite{coppola_extraordinarily_1996}.
Neural adaptation suppresses stationary floater shadows, but each eye movement resets adaptation. Because floaters lag behind eye motion, they drift across new fixation points, reintroducing motion noise \cite{silva_flow_2020}. This dynamic motivates our design goal of reducing eye movements so adaptation can persist longer, improving effective contrast and clarity on screen.

In this paper, we present three key works. First, we create a physically inspired simulation that produces video demonstrations of what floaters look like in the visual field-- to our knowledge, this is the first to introduce neural adaptation and rotational movement. Second, we use this simulation to quantitatively assess text clarity for different font types, text layouts, and floater speeds. Finally, we use these principles to build a Chrome extension that uses key bindings, RSVP, and world-in-hand navigation to minimize eye movement for both text and general UI elements. Our approach exploits neural adaptation to maximize signal to noise ratio for screen use. Our extension is freely available on the Chrome Web Store\footnote{\url{https://chromewebstore.google.com/detail/floaters-rsvp-reader/aknkkodiofapfhogikcpdoclomjlnpof}}, and we release the source code for both the extension\footnote{\url{https://github.com/jinglin-l/floaters}} and the simulation\footnote{\url{https://github.com/kenneth-ge/floater-simulation}}.

\section{Related Work}
\textbf{Simulation.} Existing mathematical models predict optical scattering and define floater opacity in the visual field \cite{harmer_optical_2022, serpetopoulos_optical_1998}, inspiring our own opacity variations. Existing video-based simulations show that floaters move in two phases, first drifting toward the position of eye movement, then settling downward \cite{noauthor_simulation_nodate-1}, mirroring first-hand observations and inspiring our two-phased motion simulation. A systematic review of floaters sketches inspired our own visual shapes and renderings \cite{lippek_investigating_2025}. Finally, studies on the gel-like non-Newtonian properties of the vitreous inspired the character of the fluid dynamics we used \cite{silva_flow_2020}. To our knowledge, our simulation is the first to: (1) incorporate neural adaptation (entoptic vanishing) for non-moving objects, (2) generate randomized videos that can be used in a computational pipeline, and (3) incorporate rotational movement and floater shape deformity.

\textbf{Computational Readability Pipeline.} We draw on existing computational pipelines that assess accessibility under low vision, like \cite{gao2025viocr}, which generated degraded renderings of text and scene content, then compared computer recognition across controlled visual conditions to approximate human reading performance under reduced acuity and contrast sensitivity. Such pipelines enable rapid, repeatable evaluation of interface design parameters without requiring large-scale user studies at early design stages. Similar end-to-end evaluation pipelines also integrate visual perturbation, layout manipulation, and automated recognition to assess readability \cite{xiong2021simulating}. These approaches treat recognition error metrics as downstream signals of pipeline performance, allowing systematic comparison of design alternatives under simulated visual noise or occlusion.

\textbf{Extension.} Existing RSVP readers such as \cite{noauthor_reedy_nodate} and \cite{noauthor_spreeder_nodate} help improve reading speed, but see reducing eye movement as a means rather than a goal. In addition, we addressed various complaints around pacing controls in existing tools \cite{treo123_software_2014}. Existing attempts to minimize user motion by moving content around the user appear frequently in VR \cite{noauthor_interaction_nodate, singleton_moving_2021, zhang_fly_nodate}, often called ``world-in-hand'', but few apply this principle to 2D navigation. Moreover, both existing RSVP tools like \cite{noauthor_swiftread_nodate} and standard browser navigation require GUI-based interactions that force gaze shifts away from content; our keyboard-first design enables control without breaking fixation. Finally, RSVP addresses only text, yet modern web interfaces are fundamentally visual. Combining RSVP with pan-based navigation provides a novel, unified approach to web navigation with minimal eye movement.

\section{Simulation and Computational Motion Analysis}
We built a 2D video simulation, showing gray translucent floater shadows atop a canvas. In addition to existing literature, our simulation was also grounded in sketch and workshop sessions in which an individual with floaters assessed their appearance against a sheer white background at 500 nits, to ensure realism.

To start, we configured the canvas and the floater appearance. We gave our canvas an aspect ratio of 3:4 to approximate the aspect ratio of the human visual field \cite{gpuguy_preferred_2012}. We made the canvas white, and the floaters a translucent gray/black, to simulate the bright backgrounds that make floaters appear most noticeable \cite{noauthor_what_nodate-1}. Each floater consists of a chain of dark joints of varying stiffness, mirroring the tangled cobweb-like shape of the floaters. We configured chain length and joint stiffness to best approximate the combination of lines, webs, and dots outlined in \cite{lippek_investigating_2025}.

In order to simulate the physics and movement, we put the floaters in a fluid field, and used XPBD \cite{macklin_xpbd_2016} to simulate movement, torque, and deformation. Since the floaters are chains with joints, this allows them to spin and deform as they do in real life, something no existing simulation we found had done before. The field first moves all floaters in some randomized eye movement direction, then moves them downward once the initial movement settles, with the velocity decaying exponentially for each phase to approximate the non-Newtonian dynamics \cite{noauthor_high-velocity_nodate} of the eye's vitreous \cite{silva_flow_2020}. The settling behavior was inspired by first-hand observation, and validated by \cite{noauthor_simulation_nodate-1}. We set the time for each phase based on stopwatch timings from an individual with floaters, resulting in an initial movement phase of $\sim$3 seconds, and a settling phase of $\sim$9 seconds.

In order to simulate neural adaptation, we made the floaters fade when their speed slows below some threshold, which we set based on the 80ms neural adaptation period \cite{coppola_extraordinarily_1996}.

Using this simulation, we generated a graph to assess how clarity in any given box in the visual field changes after eye movement. Ultimately, we found that clarity is highest at both the beginning of eye movement, and at the end. The beginning is clear because any given area of the visual field is unlikely to start already containing a floater, and the end is clear due to neural adaptation. The middle is the most noisy, as floaters repeatedly pass over all areas of the visual field.

\begin{figure}[t]
  \centering
  \includegraphics[width=0.3\linewidth]{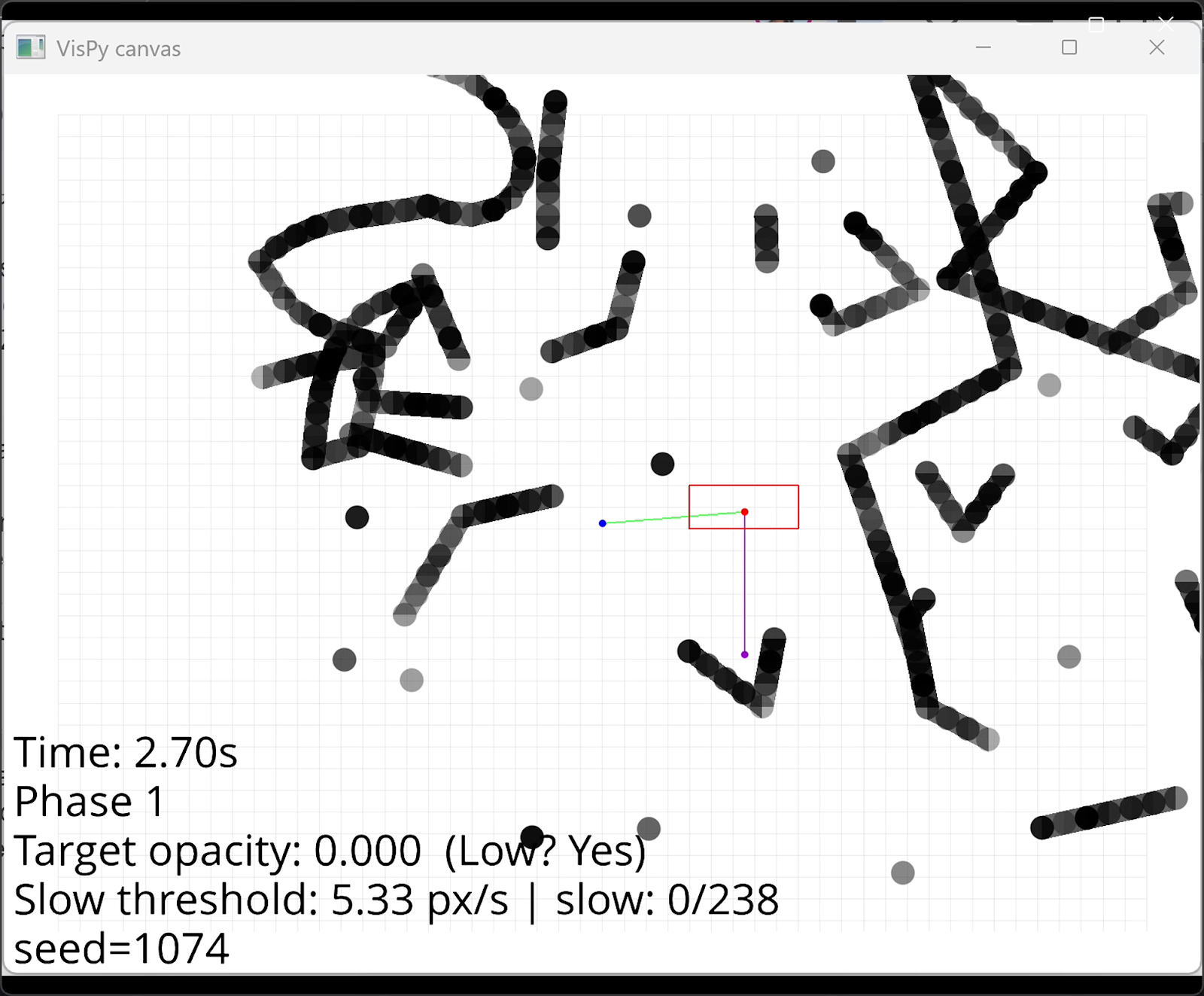}
  \hfill
  \includegraphics[width=0.55\linewidth]{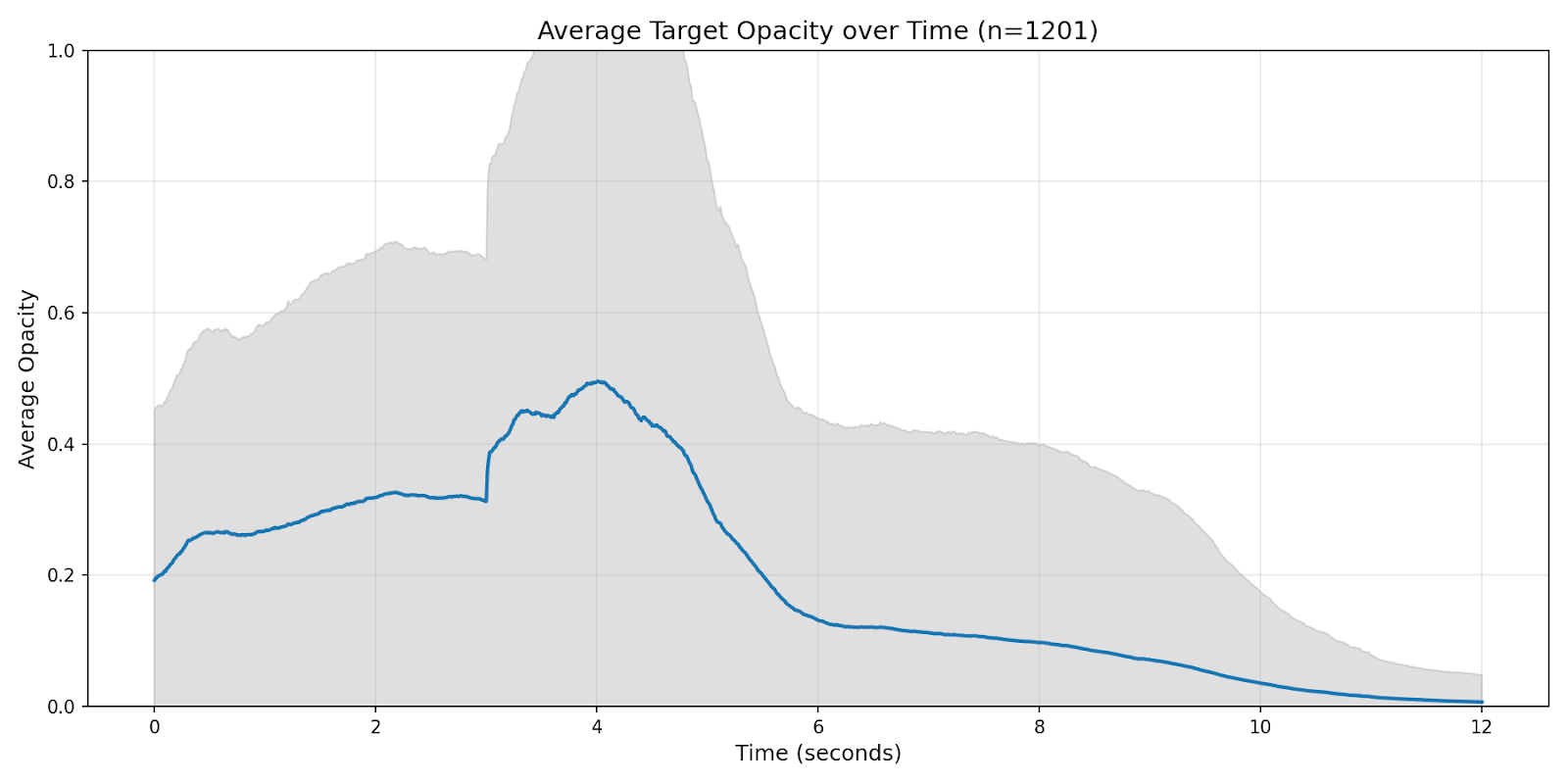}
  \caption{Simulation output (left) and resulting clarity plot (right).}
  \label{fig:simulation}
\end{figure}

\subsection{Computational Readability Pipeline}
To test the impact of simulated floaters on different font types and text layouts, we generated an overlay representing a time-averaged floater field, based on the simulated movement outlined above. Using an average allows us to approximately aggregate the effect of movement over time. To evaluate the readability of these floater overlays, we used Optical Character Recognition (OCR) as the evaluation stage of our computational readability pipeline, similar to \cite{gao2025viocr, li2025prepocr}.

We evaluated six sans-serif fonts under simulated floater occlusion using an OCR-based readability metric (results summarized in Table~\ref{tab:font_wer}). Gill Sans and Tahoma performed best, while the remaining fonts formed a closely clustered middle group. These results suggest font choice has a measurable but modest effect on text readability under floaters.

We also conducted an experiment to gauge whether reading was harder with slow floaters or fast floaters. We recorded 10 frames of the slow moving floaters and 100 frames of the fast moving floaters at a proportionally lower opacity. The slow-motion condition consistently outperformed the fast-motion condition, even when significantly darker: OCR confidence was higher for slow (0.7751) than fast (0.6602), while error rates were lower for slow---CER was 0.7591 vs 0.7734, and WER was 0.8085 vs 0.8707 for fast. All differences were statistically significant (p=0.0002).

\begin{figure}
    \centering
    \includegraphics[width=0.5\linewidth]{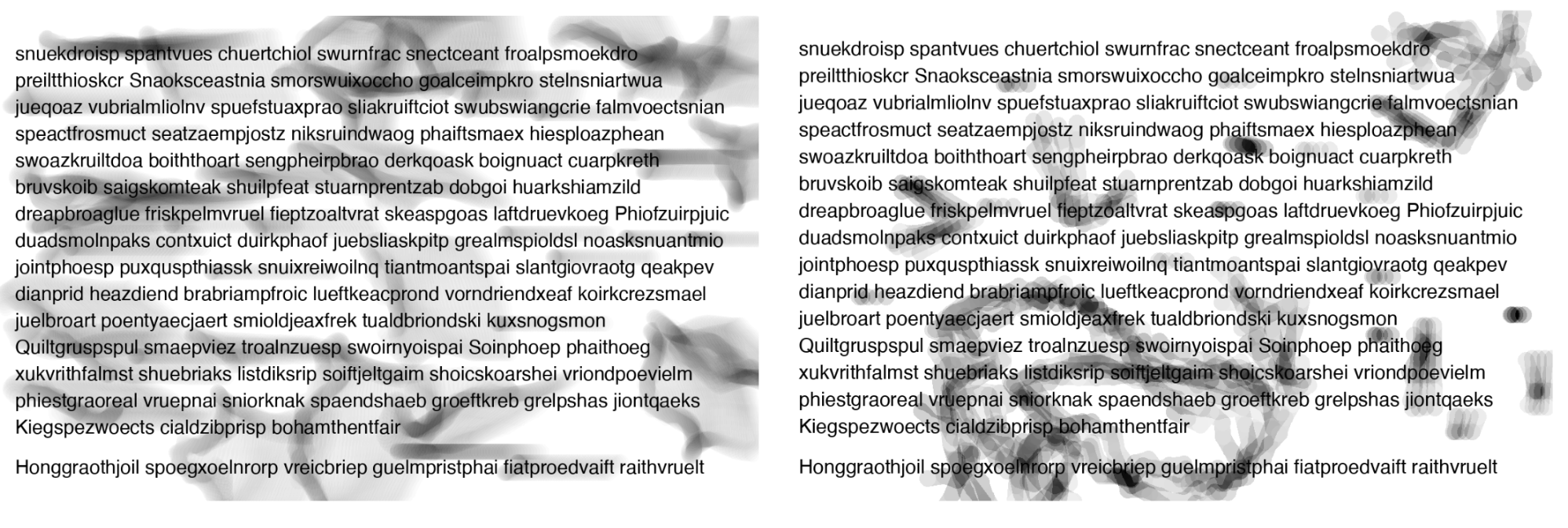}
    \caption{Fast floaters (left), compared to slow floaters (right), on randomly generated text}
    \label{fig:pipeline}
\end{figure}

We also analyzed which layouts performed optimally. Prior research \cite{galli_striving_2020} emphasized the success of single column and narrow column formats. Here, single column performed best overall while narrow single columns performed worst overall. Results are summarized in Table~\ref{tab:layout_wer}. See Fig.~\ref{fig:pipeline} for a visual example.

\begin{table}[t]
\centering
\begin{minipage}{0.48\linewidth}
\centering
\caption{OCR Word Error Rate (WER) across sans-serif fonts under simulated floater occlusion. Lower is better.}
\label{tab:font_wer}
\begin{tabular}{lc}
\toprule
\textbf{Font} & \textbf{WER} $\downarrow$ \\
\midrule
Gill Sans         & 0.877 \\
Tahoma            & 0.887 \\
Helvetica/Arial   & 0.895 \\
Helvetica Neue    & 0.894 \\
Futura            & 0.891 \\
Avenir            & 0.904 \\
\bottomrule
\end{tabular}
\end{minipage}
\hfill
\begin{minipage}{0.48\linewidth}
\centering
\caption{OCR Word Error Rate (WER) across layout conditions under floater occlusion. Lower is better.}
\label{tab:layout_wer}
\begin{tabular}{lc}
\toprule
\textbf{Layout} & \textbf{WER} $\downarrow$ \\
\midrule
Single column        & 0.895 \\
Wide spaced          & 0.915 \\
Two columns          & 0.990 \\
Narrow single column & 0.985 \\
\bottomrule
\end{tabular}
\end{minipage}
\end{table}

\section{Tool Design}

\subsection{Design Goals}

The core philosophy is to enable comfortable visual navigation of digital content by eliminating saccades (eye movements), reducing floater-induced visual disruption. Our design sessions were heavily influenced by floater physiology and gave rise to four key design goals (DGs).

\textbf{DG1: Minimize eye movements while enabling full-page visual navigation and reading.}
Our design is anchored on minimizing eye movement while preserving navigability across websites. Because neural adaptation works best when floaters stay still, every eye movement resets adaptation and reintroduces distracting motion. Therefore, our interface explicitly seeks to reduce saccades. The RSVP modal anchors gaze at a fixed focal point using ORP highlighting. Pan mode allows users to sweep across imagery and website layout by moving the viewport under a stationary crosshair.

\textbf{DG2: Minimize friction to enter, use, and exit.}
We prioritize intuitive keyboard controls for reading and cursor movements for navigation. Simple triggers and exit actions prevent excess eye movement when switching modes.

\textbf{DG3: Adapt reading pacing to cognitive load and word difficulty.}
The tool varies timing through speed controls, rewind/fast-forward, and pausing to prevent comprehension loss when encountering complex language or ideas.

\subsection{System Implementation}
\begin{figure}
    \centering
    \includegraphics[width=0.5\linewidth]{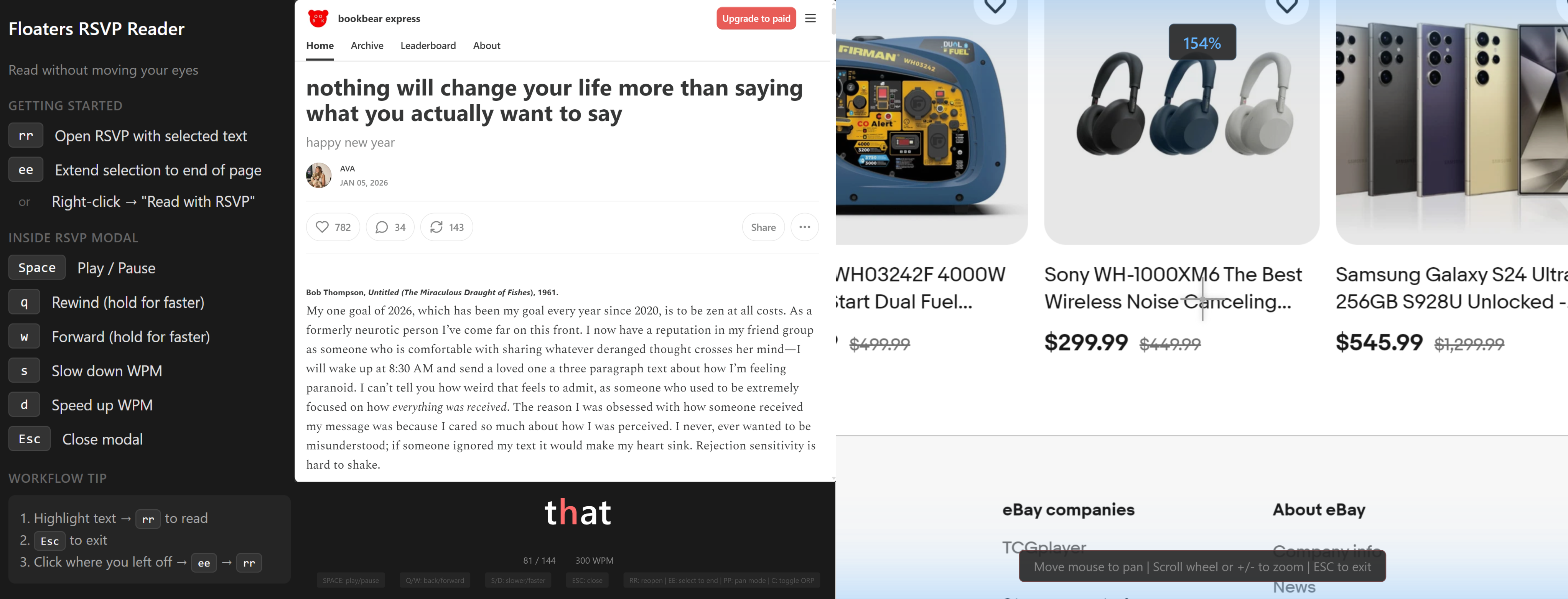}
    \caption{Extension interface, with the RSVP controls in a dropdown menu (left), RSVP display (middle), and pan mode (right)}
    \label{fig:system}
\end{figure}
In this section, we first provide a detailed explanation of end-user experience, discussed how we arrived at our final design, and finish with a technical description of system architecture. See Fig.~\ref{fig:system} for a high-level interface overview.

\subsubsection*{Overview}

\paragraph{\textbf{RSVP and Pan Mode (Addressing DG1)}}

Typically, reading text involves shifting the focus of the eye from one position to another. But, what if there was a way to do everything you would want to do on a website, without moving your eyes? Inspired by a method popular amongst many speed-reading communities, we built a Rapid Serial Visual Presentation (RSVP) reader that sequentially displays words at a fixed screen location to allow users to read a webpage without requiring eye movement. The tool is invoked by highlighting a passage of text and triggering a keyboard macro. This way, users can pick and choose what portions of a website to read.

When the modal appears, each word is displayed with a single character near its center highlighted in a distinct color and centered within the RSVP window. This anchors the word at its Optimal Recognition Point (ORP), which helps reduce the horizontal jitter that occurs when displaying words of varying lengths.

Textual content represents only one component of modern web interfaces. For non-textual elements such as photos, buttons, and general website structure, we implement a ``scene-in-hand'' panning tool (Pan mode). This interaction model draws inspiration from first-person video games, in which the user's viewport remains stationary while cursor movement manipulates the environment relative to a center point. Inside Pan mode, the cursor drags the page under a fixed crosshair. The tool also supports zoom functionality.

\paragraph{\textbf{Keybinding Design (Addressing DG2)}}

To minimize friction when entering, using, and exiting modes, all primary controls cluster on the left side of the keyboard (Q, W, S, D, Space, Esc), allowing for one handed operation of the RSVP reader while the right hand remains on the mouse for pan navigation. This layout mirrors the WASD convention used in many first-person video games. We also anticipate that over time, users will develop muscle memory of the key bindings and would be able to operate the tool without looking at the keyboard. Esc serves as an exit key for both modes.

\paragraph{\textbf{Adaptive Pacing Controls (Addressing DG3)}}

Some pieces of text are harder to understand than others. Reading comprehension and speed also vary across different people. To accommodate this variability, our RSVP reader includes speed controls: the S key slows playback and D speeds it up, letting the user slow down when encountering dense or unfamiliar content and accelerate through simpler passages. Q and W let users move the text backward and forward, supporting regression, a widely suggested way to get unstuck while reading \cite{treo123_software_2014}. Holding these keys accelerates traversal for rapid repositioning if the user wishes to go back or forward a large chunk of text, while individual presses of the key will only move the text one word per press. The spacebar pauses playback entirely, so users can digest difficult material before resuming. For personalization, we persist the speed preferences across sessions.

\subsubsection*{Architecture}

Our prototype is a Chrome extension that is focused on delivering a user friendly browser interaction. Built on Chrome's Manifest V3 framework, the extension has a background service worker, injected content scripts, and an isolated UI layer that communicates through Chrome's message-passing API. Our extension does not modify the underlying webpage.

The extension takes a target reading speed (in words per minute), converts that into exact time delays between words, and uses those delays to control how quickly words are shown. English words vary in length and as a result a horizontal jitter occurs within the RSVP display itself. To address this, we implement the Optimal Recognition Point (ORP) algorithm, which anchors each word at its perceptual center, which shifts inward as word length increases.

Pan mode was the trickier feature to get right. Our first attempt used native scroll APIs (scrollBy), but sites with custom scroll containers or virtual scrolling behaved unpredictably, causing stuttering and weird jumps on some pages. Ultimately, we decided against scroll entirely and applied translate3d transforms directly to document.body instead. This gives us smooth 60fps panning on any site, regardless of how it handles scrolling internally.

To capture raw mouse deltas, we use the Pointer Lock API, which provides movementX and movementY values representing true physical mouse displacement. Mapping these deltas directly to page movement often results in a navigation that feels jittery, so instead, we feed them into an interpolation loop. The page glides toward a target position each frame rather than snapping to it, producing panning that naturally decelerates. Broadly, getting the feel right required some iteration. For example, our initial zoom step (8\% per scroll tick) made the page bounce around erratically, and we dialed it down to 3\%. Pan speed also needed to scale with zoom level. For instance, at 2x zoom, the same mouse movement covers twice the visual distance, so without compensation the page moves too fast.

We wanted entering and exiting modes to feel deliberate and intuitive, reducing friction to use them. We considered single keypresses and modifier combos like Ctrl+Shift+R, but ultimately found that they were awkward, and conflicted in unpredictable ways with browser shortcuts. In the end, we landed on double-tap activation: tap `r' twice quickly for RSVP, `p' twice for pan mode. A 400ms window distinguishes intentional double-taps from coincidental repeated letters.

\section{Future Work}
Realistic physics simulation and randomization created tradeoffs with respect to realism of floater shape. Our work approximates the macro behavior, but future work will incorporate human-drawn floaters to further improve realism. Additionally, we would like to generalize our Computational Readability Pipeline to solve problems beyond the scope of floaters. We have launched our floater browser extension and will conduct a follow up user study.

\section*{Acknowledgements}
This work was self-funded. We are grateful for the opportunity to work on this problem.


\begin{thebibliography}{99}

\bibitem{admin_donna_2021}
Admin.
\newblock Donna {M}. {\textbar} Floater Stories, June 2021.
\newblock \url{https://floaterstories.com/donna-m/}. Accessed: 2026-05-12.

\bibitem{charm_high_2013}
J.~Charm and P.~Cho.
\newblock High myopia--partial reduction ortho-k: A 2-year randomized study.
\newblock {\em Optometry and Vision Science}, 90(6):530, 2013.

\bibitem{coppola_extraordinarily_1996}
D.~Coppola and D.~Purves.
\newblock The extraordinarily rapid disappearance of entoptic images.
\newblock {\em Proceedings of the National Academy of Sciences}, 93(15):8001--8004, 1996.

\bibitem{noauthor_efficacy_nodate}
Efficacy and safety of early {YAG} laser vitreolysis for symptomatic vitreous floaters: the study protocol for a randomized clinical trial, 2024.
\newblock \url{https://link.springer.com/article/10.1186/s13063-024-07924-1}. Accessed: 2026-05-12.

\bibitem{noauthor_features_nodate}
D.~Fiset, C.~Blais, C.~{\'E}thier-Majcher, M.~Arguin, D.~Bub, and F.~Gosselin.
\newblock Features for identification of uppercase and lowercase letters.
\newblock {\em Psychological Science}, 2008.

\bibitem{galli_striving_2020}
C.~Galli, M.~T. Colangelo, and S.~Guizzardi.
\newblock Striving for modernity: Layout and abstracts in the biomedical literature.
\newblock {\em Publications}, 8(3):38, 2020.

\bibitem{gao2025viocr}
Q.~Gao, R.~Manduchi, P.~Y. Ramulu, G.~E. Legge, and Y.~Xiong.
\newblock {VI-OCR}: Visually impaired optical character recognition pipeline for text accessibility assessment.
\newblock {\em Scientific Reports}, 15:30982, 2025.

\bibitem{gouliopoulos_association_2024}
N.~Gouliopoulos, D.~Oikonomou, F.~Karygianni, A.~Rouvas, S.~Kympouropoulos, and M.~M. Moschos.
\newblock The association of symptomatic vitreous floaters with depression and anxiety.
\newblock {\em International Ophthalmology}, 44(1):218, 2024.

\bibitem{gpuguy_preferred_2012}
gpuguy.
\newblock Preferred aspect ratio for human eyes.
\newblock Biology Stack Exchange, November 2012.
\newblock \url{https://biology.stackexchange.com/q/5128}. Accessed: 2026-05-12.

\bibitem{harmer_optical_2022}
S.~W. Harmer, A.~J. Luff, and G.~Gini.
\newblock Optical scattering from vitreous floaters.
\newblock {\em Bioelectromagnetics}, 43(2):90--105, 2022.

\bibitem{noauthor_high-velocity_nodate}
High-velocity impact of solid objects on non-{N}ewtonian fluids.
\newblock {\em Scientific Reports}, 2019.
\newblock \url{https://www.nature.com/articles/s41598-018-37543-1}. Accessed: 2026-05-12.

\bibitem{noauthor_ive_nodate}
I've been eating pineapple in the last few weeks to see if it gets rid of my eye floaters.
\newblock Hacker News, 2021.
\newblock \url{https://news.ycombinator.com/item?id=29460531}. Accessed: 2026-05-12.

\bibitem{noauthor_interaction_nodate}
Interaction metaphor --- an overview.
\newblock ScienceDirect Topics, 2013.
\newblock \url{https://www.sciencedirect.com/topics/computer-science/interaction-metaphor}. Accessed: 2026-05-12.

\bibitem{katsanos_safety_2020}
A.~Katsanos, N.~Tsaldari, K.~Gorgoli, F.~Lalos, M.~Stefaniotou, and I.~Asproudis.
\newblock Safety and efficacy of {YAG} laser vitreolysis for the treatment of vitreous floaters: An overview.
\newblock {\em Advances in Therapy}, 37(4):1319--1327, 2020.

\bibitem{legge_psychophysics_1985}
G.~E. Legge, D.~G. Pelli, G.~S. Rubin, and M.~M. Schleske.
\newblock Psychophysics of reading---{I}. Normal vision.
\newblock {\em Vision Research}, 25(2):239--252, 1985.

\bibitem{levi_crowding--essential_2008}
D.~M. Levi.
\newblock Crowding---an essential bottleneck for object recognition: A mini-review.
\newblock {\em Vision Research}, 48(5):635--654, 2008.

\bibitem{li2025prepocr}
Y.~Li, R.~Zhang, H.~Chen, and W.~Wang.
\newblock {PreP-OCR}: Document image restoration and {OCR} in a unified pipeline.
\newblock {\em arXiv preprint arXiv:2505.20429}, 2025.

\bibitem{lippek_investigating_2025}
J.~Lippek, L.~Rynko, C.~Framme, O.~Kermani, S.~Johannsmeier, and T.~Ripken.
\newblock Investigating symptomatic vitreous opacities: An online survey and field of view reconstruction.
\newblock {\em Klinische Monatsbl{\"a}tter f{\"u}r Augenheilkunde}, 242(10):991--1000, 2025.

\bibitem{macklin_xpbd_2016}
M.~Macklin, M.~M{\"u}ller, and N.~Chentanez.
\newblock {XPBD}: Position-based simulation of compliant constrained dynamics.
\newblock In {\em Proceedings of the 9th International Conference on Motion in Games}, pages 49--54, 2016.

\bibitem{pelli_crowding_2007}
D.~G. Pelli, K.~A. Tillman, J.~Freeman, M.~Su, T.~D. Berger, and N.~J. Majaj.
\newblock Crowding and eccentricity determine reading rate.
\newblock {\em Journal of Vision}, 7(2):20.1--36, 2007.

\bibitem{noauthor_reedy_nodate}
Reedy. Intelligent reader, 2024.
\newblock \url{https://reedy-reader.com/}. Accessed: 2026-05-12.

\bibitem{sawides_vision_2011}
L.~Sawides, P.~de~Gracia, C.~Dorronsoro, M.~A. Webster, and S.~Marcos.
\newblock Vision is adapted to the natural level of blur present in the retinal image.
\newblock {\em PLOS ONE}, 6(11):e27031, 2011.

\bibitem{sebag_vitrectomy_2014}
J.~Sebag, K.~M.~P. Yee, C.~A. Wa, L.~C. Huang, and A.~A. Sadun.
\newblock Vitrectomy for floaters: Prospective efficacy analyses and retrospective safety profile.
\newblock {\em RETINA}, 34(6):1062, 2014.

\bibitem{sebag_vitreous_2020}
J.~Sebag.
\newblock Vitreous and vision degrading myodesopsia.
\newblock {\em Progress in Retinal and Eye Research}, 79:100847, 2020.

\bibitem{serpetopoulos_optical_1998}
C.~N. Serpetopoulos and R.~A. Korakitis.
\newblock An optical explanation of the entoptic phenomenon of `clouds' in posterior vitreous detachment.
\newblock {\em Ophthalmic \& Physiological Optics}, 18(5):446--451, 1998.

\bibitem{silva_flow_2020}
A.~F. Silva, F.~Pimenta, M.~A. Alves, and M.~S.~N. Oliveira.
\newblock Flow dynamics of vitreous humour during saccadic eye movements.
\newblock {\em Journal of the Mechanical Behavior of Biomedical Materials}, 110:103860, 2020.

\bibitem{noauthor_simulation_nodate-1}
Simulation of floaters.
\newblock JN Learning, AMA Ed Hub, 2019.
\newblock \url{https://edhub.ama-assn.org/jn-learning/video-player/2522137}. Accessed: 2026-05-12.

\bibitem{singleton_moving_2021}
A.~Singleton.
\newblock Moving around in {VR}: Drag world.
\newblock PintSizedRobotNinja, Medium, July 2021.
\newblock \url{https://medium.com/pintsizedrobotninja/moving-around-in-vr-drag-world-edbfa6786f8c}. Accessed: 2026-05-12.

\bibitem{noauthor_spreeder_nodate}
Spreeder --- speed reading app \& software, 2024.
\newblock \url{https://www.spreeder.com/}. Accessed: 2026-05-12.

\bibitem{noauthor_swiftread_nodate}
{SwiftRead} --- speed reading software, 2024.
\newblock \url{https://swiftread.com/}. Accessed: 2026-05-12.

\bibitem{the_floater_doctor__james_h_johnson_md_vitreous_2020}
The Floater Doctor (James H. Johnson MD).
\newblock Vitreous eye floater destruction \& relief without surgery: Example \& optics of treatment.
\newblock YouTube, February 2020.
\newblock \url{https://www.youtube.com/watch?v=pjGa8qOZvLE}. Accessed: 2026-05-12.

\bibitem{treo123_software_2014}
Treo123.
\newblock Software that speeds up your reading to 500 words per minute.
\newblock Reddit r/books, February 2014.
\newblock \url{https://www.reddit.com/r/books/comments/1yvvam/software_that_speeds_up_your_reading_to_500_words/}. Accessed: 2026-05-12.

\bibitem{noauthor_vitreous_nodate}
Vitreous opacity --- an overview.
\newblock ScienceDirect Topics, 2024.
\newblock \url{https://www.sciencedirect.com/topics/medicine-and-dentistry/vitreous-opacity}. Accessed: 2026-05-12.

\bibitem{wagemans_century_2012}
J.~Wagemans, J.~H. Elder, M.~Kubovy, S.~E. Palmer, M.~A. Peterson, M.~Singh, and R.~von~der Heydt.
\newblock A century of {Gestalt} psychology in visual perception {I}: Perceptual grouping and figure-ground organization.
\newblock {\em Psychological Bulletin}, 138(6):1172--1217, 2012.

\bibitem{wagle_utility_2011}
A.~M. Wagle, W.-Y. Lim, T.-P. Yap, K.~Neelam, and K.-G. Au~Eong.
\newblock Utility values associated with vitreous floaters.
\newblock {\em American Journal of Ophthalmology}, 152(1):60--65.e1, 2011.

\bibitem{webb_prevalence_2013}
B.~F. Webb, J.~R. Webb, M.~C. Schroeder, and C.~S. North.
\newblock Prevalence of vitreous floaters in a community sample of smartphone users.
\newblock {\em International Journal of Ophthalmology}, 6(3):402--405, 2013.

\bibitem{webster_neural_2002}
M.~A. Webster, M.~A. Georgeson, and S.~M. Webster.
\newblock Neural adjustments to image blur.
\newblock {\em Nature Neuroscience}, 5(9):839--840, 2002.

\bibitem{noauthor_what_nodate}
What are eye floaters?
\newblock Cleveland Clinic, 2023.
\newblock \url{https://my.clevelandclinic.org/health/symptoms/14209-eye-floaters-myodesopias}. Accessed: 2026-05-12.

\bibitem{noauthor_what_nodate-1}
What causes floaters \& flashers in the eye.
\newblock EyeCare Associates, 2025.
\newblock \url{https://www.webeca.com/eye-care-resources/eye-health/what-causes-floaters-and-flashes-in-the-eye}. Accessed: 2026-05-12.

\bibitem{woudstra_de_jong_impact_2025}
J.~E. Woudstra-de Jong, S.~S. Manning-Charalampidou, J.~H.~R. Vingerling, S.~J.~F. Gerbrandy, K.~Pesudovs, and J.~J. Busschbach.
\newblock The impact of vitreous floaters on quality of life: A qualitative study.
\newblock {\em Journal of Patient-Reported Outcomes}, 9(1):102, 2025.

\bibitem{xiong2021simulating}
Y.~Xiong, Q.~Lei, A.~Calabr{\`e}se, and G.~E. Legge.
\newblock Simulating visibility and reading performance in low vision.
\newblock {\em Frontiers in Neuroscience}, 15:671121, 2021.

\bibitem{zhang_fly_nodate}
F.~Zhang, J.~S. Zhu, E.~Lank, K.~Katsuragawa, and J.~Zhao.
\newblock Fly the moon to me: Bimanual {3D} locomotion in virtual reality by manipulating the position of the destination object.
\newblock {\em Proceedings of Graphics Interface}, 2023.

\end{thebibliography}
\end{document}